\documentstyle[12pt]{article}
\newcommand{\bra}{\langle}
\newcommand{\ket}{\rangle}
\newcommand{\R}{\mbox{\boldmath $ R $}}
\newcommand{\Z}{\mbox{\boldmath $ Z $}}
\newcommand{\map}{\mbox{Map}}
\newcommand{\diff}{\mbox{Diff}}

\begin{document}
\baselineskip 5mm
\begin{flushright}
{\small DPNU-95-06}
\\
{\small %
hep-th/9502159
}
\end{flushright}
\baselineskip 8mm
\begin{center}
{\Large \bf %
Topology and Inequivalent Quantizations \\
of Abelian Sigma Model}\footnote{%
Presented on 19 January 1995
at Second Pasific Winter School for Theoretical Physics
held at Sorak Mountain in Korea.
At that time the title of seminar was
``Winding number, zero-mode and commutators
of abelian sigma model in (1+1) dimensions'',
but it is changed here.}
\vspace{1mm} \\
{\large Shogo Tanimura}\footnote{%
e-mail address : tanimura@eken.phys.nagoya-u.ac.jp}
\baselineskip 6mm
\\
{\it Department of Physics, Nagoya University,
Nagoya 464-01, Japan}\footnote{%
Leaving for Department of Physics, University of Tokyo,
Bunkyo-ku,Tokyo 113, Japan}
\vspace{3mm}
\\
\begin{minipage}[t]{120mm}
\baselineskip 12pt
{\small
	The abelian sigma model in (1+1) dimensions
	is a field theoretical model which has a field
	$ \phi : S^1 \to S^1 $.
	An algebra of the quantum field is defined
	respecting the topological aspect of the model.
	It is shown that
	the zero-mode has an infinite number of inequivalent quantizations.
	It is also shown that
	when a central extension is introduced into the algebra,
	the winding operator and the momenta operators satisfy
	anomalous commutators.
}
\end{minipage}
\end{center}
%
\baselineskip 15pt
\noindent
{\bf 1. Introduction}
\vspace{0.1cm} \\
In both field theory and string theory
there are several models which have manifold-valued variables,
for instance, the nonlinear sigma model and
the toroidal compactification model of closed bosonic string.
To study global aspects of these models in quantum theory,
we should have a quantization scheme respecting topological nature.
However in the usual scheme of canonical quantization
and perturbation method,
the global aspects are obscure.
On the other hand it is known~[1], [2] that
there exist an infinite number of inequivalent quantizations
on a topologically nontrivial manifold
even when it is a finite-dimensional manifold.
In this note we consider the abelian sigma model in (1+1) dimensions
to explore a system having infinite degrees of freedom.
The model has a manifold-valued field $ \phi : S^1 \to S^1 $.
 For detail see the reference~[3].
%
%
%
\vspace{0.2cm} \\
{\bf 2. Ohnuki-Kitakado representation}
\vspace{0.1cm} \\
We briefly review quantum mechanics of a particle on $ S^1 $~[1].
Let us assume that
a unitary operator $ \hat{U} $ and a self-adjoint operator $ \hat{P} $
satisfy the commutation relation
\begin{equation}
	[ \, \hat{P} \, , \, \hat{U} \, ] = \hat{U}.
	\label{2.1}
\end{equation}
An irreducible representation of the above algebra is called
quantum mechanics on $ S^1 $.
The operators $ \hat{U} $ and $ \hat{P} $ are called
a position operator and a momentum operator, respectively~[4].
\par
A representation space is constructed as follows.
Let $ | \alpha \ket $ be an eigenvector of $ \hat{P} $
with a real eigenvalue $ \alpha $;
$ \hat{P} \, | \alpha \ket = \alpha \, | \alpha \ket $.
Assume that $ \bra \alpha | \alpha \ket = 1 $.
The commutator (\ref{2.1}) implies that
the operator $ \hat{U} $ increases an eigenvalue of $ \hat{P} $ by a unit.
If we put
\begin{equation}
	| n + \alpha \ket := \hat{U}^n \, | \alpha \ket
	\;\;\;
	( n = 0 , \pm 1 , \pm 2 , \cdots ),
	\label{2.2}
\end{equation}
it is easily seen that
\begin{equation}
	\hat{P} \, | n + \alpha \ket = ( n + \alpha ) \, | n + \alpha \ket,
	\;\;\;
	\hat{U} \, | n + \alpha \ket = | n + 1 + \alpha \ket.
	\label{2.4}
\end{equation}
Unitarity of $ \hat{U} $ and self-adjointness of $ \hat{P} $ imply that
\begin{equation}
	\bra m + \alpha | n + \alpha \ket = \delta_{ m \, n }.
	\label{2.5}
\end{equation}
Let $ H_\alpha $ denote the Hilbert space defined by completing
the space of finite linear combinations of
$ | n + \alpha \ket \, ( n = 0 , \pm 1 , \pm 2 , \cdots ) $.
By (\ref{2.4}), $ H_\alpha $ becomes
an irreducible representation space of the algebra (\ref{2.1}).
$ H_\alpha $ and $ H_\beta $ are unitary equivalent
if and only if the difference $ ( \alpha - \beta ) $ is an integer.
Consequently there exists an inequivalent representation
for each value of the parameter $ \alpha $ ranging over
$ 0 \le \alpha < 1 $.
Physical significance of $ \alpha $ is further discussed
in the reference~[4].
%
%
%
\vspace{0.2cm} \\
{\bf 3. Algebra}
\vspace{0.1cm} \\
Next we shall propose an algebra of the sigma model.
In the context of classical theory,
the model has a field variable $ \phi \in Q = \map(S^1; S^1) $.
Let $ \Gamma = \map(S^1; U(1)) $ a group by pointwise composition.
The group $ \Gamma $ acts on the configuration space $ Q $
by pointwise action of $ U(1) $ on $ S^1 $.
To quantize this system let us assume that
$ \hat{\phi}(\theta) $ is a unitary operator for each point
$ \theta \in S^1 $ and
$ \hat{V}(\gamma) $ is a unitary operator for each element
$ \gamma \in \Gamma $.
Moreover we assume the following algebra
\begin{eqnarray}
	&&
	\hat{\phi}(\theta)  \, \hat{\phi}(\theta') =
	\hat{\phi}(\theta') \, \hat{\phi}(\theta),
	\label{3.8}
	\\
	&&
	\hat{V}(\gamma)^\dagger \, \hat{\phi}(\theta) \, \hat{V}(\gamma) =
	\gamma(\theta) \, \hat{\phi}(\theta),
	\label{3.9}
	\\
	&&
	\hat{V}(\gamma_1) \, \hat{V}(\gamma_2) =
	e^{ - i \, c ( \gamma_1, \gamma_2 ) } \,
	\hat{V}(\gamma_1 \cdot \gamma_2)
	\;\;\;
	( \gamma_1, \, \gamma_2 \in \Gamma ).
	\label{3.10}
\end{eqnarray}
In the last line a function $ c : \Gamma \times \Gamma \to \R $
is called a central extension, which satisfies the cocycle condition;
$
	  c( \gamma_1 , \gamma_2 )
	+ c( \gamma_1 \gamma_2 , \gamma_3 )
	= c( \gamma_1 , \gamma_2 \gamma_3 )
	+ c( \gamma_2 , \gamma_3 )
$
$ ( \mbox{mod} \: 2 \pi ) $.
We call the algebra (\ref{3.8})-(\ref{3.10})
the fundamental algebra of the abelian sigma model.
To consider a nontrivial central extension we decompose $ \gamma $ as
\begin{equation}
	\gamma ( \theta ) = e^{ i \, ( \mu + f(\theta) + m \, \theta ) },
	\label{3.14}
\end{equation}
where $ \mu \in \R $,
$ f \in
\map_0 ( S^1 ; \R ) :=
\{ g : S^1 \to \R \, | \,
C^\infty , \, \int_0^{2 \pi} g(\theta) \, d \theta = 0 \} $,
and $ m \in \Z $.
 For $ \gamma_1 $ shown at (\ref{3.14}) and
$ \gamma_2 ( \theta )
= \exp[ i \, ( \mu_2 + f_2(\theta) + m_2 \, \theta ) ] $,
we define
\begin{equation}
	c( \gamma_1 , \gamma_2 ) :=
	k
	\left\{
	  m_1 \, \mu_2
	- \mu_1 \, m_2
	+
	\frac{1}{4 \pi} \int_0^{2 \pi}
	\left(
	f_1'(\theta) f_2(\theta) - f_1(\theta) f_2'(\theta)
	\right)
	d \theta
	\right\},
	\label{3.19}
\end{equation}
where $ k $ is a non-zero integer.
This central extension is the simplest but nontrivial one
which is invariant under the action of $ \diff^+( S^1 ) $;
$ c( \gamma_1 \circ \omega , \gamma_2 \circ \omega )
= c( \gamma_1 , \gamma_2 ) $
for any $ \omega \in \diff^+( S^1 ) $.
The group $ \Gamma $ associated with such an invariant central extension
is called a Kac-Moody group~[5].
%
%
\par
{\it a. Algebra without central extension :}
To clarify geometrical implication of the algebra
we shall decompose the degrees of freedom of
$ \hat{\phi} $ and $ \hat{V} $.
At first we put $ c \equiv 0 $.
We introduce
a unitary operator $ \hat{U} $,
self-adjoint operators\footnote{%
Expressing rigorously $ \hat{\varphi} (\theta) $
is an operator-valued distribution.
$ \exp( i \hat{\varphi} (\theta) ) $ must be regularized
by the normal product procedure[3].}
$ \hat{\varphi} ( \theta ) $ for each $ \theta \in S^1 $
constrained by
$ \int_0^{ 2 \pi } \hat{\varphi} ( \theta ) \, d \theta = 0 $,
and a self-adjoint operator $ \hat{N} $ satisfying
$ \exp( 2 \pi i \, \hat{N} ) = 1 $,
which is called the integer condition for $ \hat{N} $.
We demand that the quantum field $ \hat{\phi} ( \theta ) $ is expressed
by them as
\begin{equation}
	\hat{\phi} ( \theta )
	=
	\hat{U} \,
	e^{ i \, ( \hat{\varphi} ( \theta ) + \hat{N} \theta ) }.
	\label{3.22}
\end{equation}
Geometrical meaning of this decomposition is apparent;
$ \hat{U} $ describes the zero-mode or collective motion
of the field $ \hat{\phi} $;
$ \hat{\varphi} $ describes fluctuation or local degrees of freedom;
$ \hat{N} $ is nothing but the winding number.
Topologically nontrivial parts are $ \hat{U} $ and $ \hat{N} $.
Next, corresponding to (\ref{3.14}) we introduce
a self-adjoint operator $ \hat{P} $,
self-adjoint operators $ \hat{\pi} ( \theta ) $
for each $ \theta \in S^1 $ constrained by
$ \int_0^{ 2 \pi } \hat{\pi} ( \theta ) \, d \theta = 0 $,
and a unitary operator $ \hat{W} $.
When $ \gamma $ is given by (\ref{3.14}), we define $ \hat{V} ( \gamma ) $
by
\begin{equation}
	\hat{V} ( \gamma )
	=
	e^{ - i \, \mu \, \hat{P} } \,
	\exp
	\left[
	- i \int_0^{2 \pi} f( \theta ) \, \hat{\pi} ( \theta )
	\, d \theta
	\right]
	\hat{W}^m.
	\label{3.24}
\end{equation}
Using these operators the fundamental algebra is now rewritten as
\begin{eqnarray}
	&&
	[ \, \hat{P} , \hat{U} \, ] = \hat{U},
	\label{3.25}
	\\
	&&
	[ \, \hat{\varphi} ( \theta ) , \hat{\pi} ( \theta' ) \, ]
	=
	i \Bigl( \delta( \theta - \theta' ) - \frac{1}{2 \pi} \Bigr),
	\label{3.26}
	\\
	&&
	[ \, \hat{N} , \hat{W} \, ] = \hat{W},
	\label{3.27}
\end{eqnarray}
and all other commutators vanish.
In (\ref{3.26}) the $ \delta $-function is periodic
with the periodicity $ 2 \pi $.
Observing the relation (\ref{3.27}) we call
$ \hat{W} $ the winding operator.
%
%
\par
{\it b. Algebra with central extension :}
If the central extension (\ref{3.19}) is included,
the decomposition (\ref{3.24}) of $ \hat{V} $
should be modified a little.
It is replaced by
\begin{equation}
	\hat{V} ( \gamma )
	=
	e^{ - i k m \mu }
	\,
	e^{ - i \mu \hat{P} }
	\exp
	\left[
		- i
		\int_0^{2 \pi} f( \theta ) \, \hat{\pi}(\theta )
		\, d \theta
	\right]
	\hat{W}^m.
	\label{3.37}
\end{equation}
 Furthermore the following commutation relations should be added
to (\ref{3.25})-(\ref{3.27});
\begin{eqnarray}
	&&
	[ \, \hat{P} , \hat{W} \, ] = - 2 k \, \hat{W},
	\label{3.34}
	\\
	&&
	[ \, \hat{\pi} (\theta) , \hat{\pi} (\theta') \, ]
	=
	- \, \frac{i k}{\pi} \, \delta' ( \theta - \theta' ).
	\label{3.35}
\end{eqnarray}
(\ref{3.34}) says that
the zero-mode momentum $ \hat{P} $ is decreased by $ 2 k $ units
when the winding number $ \hat{N} $ is increased by one unit
under the operation of $ \hat{W} $.
This is an inevitable consequence of the central extension.
We call this phenomenon ``twist''.
%
%
%
\vspace{0.2cm} \\
{\bf 4. Representations}
\vspace{0.1cm} \\
%
%
{\it a. Without the central extension :}
Representations of the algebra
defined by (\ref{3.25})-(\ref{3.27}) and other vanishing commutators
are constructed below.
$ \hat{P} $ and $ \hat{N} $ are self-adjoint
and that $ \hat{U} $ and $ \hat{W} $ are unitary.
Both of the relations (\ref{3.25}) and (\ref{3.27})
are isomorphic to (\ref{2.1}).
Hence the Ohnuki-Kitakado representation can be used for them.
$ \hat{P} $ and $ \hat{U} $ act on the Hilbert space $ H_\alpha $
via (\ref{2.4}).
$ \hat{N} $ and $ \hat{W} $ act on another Hilbert space $ H_\beta $ via
$ \hat{N} \, | n + \beta \ket = ( n + \beta ) \, | n + \beta \ket $ and
$ \hat{W} \, | n + \beta \ket = | n + 1 + \beta \ket $.
The value of $ \alpha $ is arbitrary.
However $ \beta $ is restricted to be an integer
if we impose the integer condition $ \exp ( 2 \pi i \, \hat{N} ) = 1 $.
 For $ \hat{\varphi} $ and $ \hat{\pi} $
we define operators $ \hat{a}_n $ and $ \hat{a}_n^\dagger $ by
\begin{equation}
	\hat{\varphi} (\theta)
	=
	\frac{1}{2 \pi} \, \sum_{ n \ne 0 } \,
	\sqrt{ \frac{ \pi }{ | n | } } \,
	( \hat{a}_n         \, e^{   i \, n \, \theta }
	+ \hat{a}_n^\dagger \, e^{ - i \, n \, \theta } ),
	\;\;
	\hat{\pi} (\theta)
	=
	\frac{i}{2 \pi} \, \sum_{ n \ne 0 } \,
	\sqrt{ \pi | n | } \,
	( - \hat{a}_n         \, e^{   i \, n \, \theta }
	  + \hat{a}_n^\dagger \, e^{ - i \, n \, \theta } ).
	\label{4.4}
\end{equation}
It is easily verified that the commutator (\ref{3.26}) is equivalent to
\begin{equation}
	[ \, \hat{a}_m , \hat{a}_n^\dagger ] = \delta_{ m \, n }
	\;\;\;
	( m , n = \pm 1 , \pm 2 , \cdots )
	\label{4.5}
\end{equation}
with the other vanishing commutators.
Hence the ordinary Fock space $ F $ gives a representation
of $ \hat{a} $'s and $ \hat{a}^\dagger $'s.
Consequently
the tensor product space $ H_\alpha \otimes H_0 \otimes F $
$ (0 \le \alpha < 1) $ gives an irreducible representation
of the algebra (\ref{3.25})-(\ref{3.27}).
%
%
\par
{\it b. With the central extension :}
Representations of the algebra
defined by
(\ref{3.25}), (\ref{3.26}), (\ref{3.27}), (\ref{3.34}) and (\ref{3.35})
and other vanishing commutators
are also constructed in a similar way.
The twist relation (\ref{3.34}) is represented by
\begin{equation}
	\left\{
	\begin{array}{ll}
	\hat{P}          \, | \, p + \alpha ; \, n \ket
	= ( p + \alpha ) \, | \, p + \alpha ; \, n \ket,
	&
	\hat{U} \, | \, p     + \alpha ; \, n \ket
	=          | \, p + 1 + \alpha ; \, n \ket,
	\\
	\hat{N} \, | \, p + \alpha ; \, n \ket
	=    n     | \, p + \alpha ; \, n \ket,
	&
	\hat{W} \, | \, p       + \alpha ; \, n     \ket
	=          | \, p - 2 k + \alpha ; \, n + 1 \ket.
	\end{array}
	\right.
	\label{4.11}
\end{equation}
The Hilbert space formed by completing the space of linear combinations of
$ | \, p + \alpha ; \, n \ket $
is denoted by $ T_{k \alpha} $.
($ T $ indicates ``twist''.)
Taking the anomalous commutator (\ref{3.35}) into account
the Fourier expansions (\ref{4.4})
of $ \hat{\varphi} $ and $ \hat{\pi} $ are changed to
\begin{eqnarray}
	&&
	\hat{\varphi} (\theta)
	=
	\sum_{ n \ne 0 } \,
	\frac{1}{ \sqrt{ 2 | k n | } } \,
	( \hat{a}_n         \, e^{   i \, n \, \theta }
	+ \hat{a}_n^\dagger \, e^{ - i \, n \, \theta } ),
	\nonumber \\
	&&
	\hat{\pi} (\theta)
	=
	\frac{i}{2 \pi} \, \sum_{ n \ne 0 } \,
	\theta( k n ) \,
	\sqrt{ 2 | k n | } \,
	( - \hat{a}_n         \, e^{   i \, n \, \theta }
	  + \hat{a}_n^\dagger \, e^{ - i \, n \, \theta } ).
	\label{4.14}
\end{eqnarray}
It should be noticed that
only positive $ n $'s appear in the expansion of $ \hat{\pi} $
even though both of positive and negative $ n $'s appear in
$ \hat{\varphi} $.
Thus the tensor product space $ T_{k \alpha} \otimes F $
$ (0 \le \alpha < 1) $
gives an irreducible representation of the fundamental algebra.
%
%
%
\vspace{0.2cm} \\
{\bf Acknowledgment}
\vspace{0.1cm} \\
I thank deeply Yoonbai Kim,
who encouraged me to attend the winter school.
%
%
%
\vspace{0.2cm} \\
{\bf References}
\vspace{0.1cm}
\par
[1]
Y. Ohnuki and S. Kitakado,
{\it J. Math. Phys.} {\bf 34} (1993) 2827.
\par
[2]
G. W. Mackey,
{\it Induced Representations of Groups and Quantum Mechanics},
(W. A. Benjamin, INC., New York, 1968);
N. P. Landsman and N. Linden,
{\it Nucl. Phys.} {\bf B365} (1991) 121;
S. Tanimura,
Nagoya University preprint DPNU-93-21 (1993), hep-th/9306144.
\par
[3]
S. Tanimura,
{\it Phys. Lett.} {\bf 340B} (1994) 57;
Nagoya University preprint DPNU-94-60 (1994),
hep-th/9412174
(submitted to {\it Nucl. Phys.}).
\par
[4]
S. Tanimura,
{\it Prog. Theor. Phys.} {\bf 90} (1993) 271.
\par
[5]
G. Segal,
{\it Commun. Math. Phys.} {\bf 80} (1981) 301.
\end{document}